\documentclass[twocolumn,showpacs,amsmath,amssymb,prl,superscriptaddress,10pt]{revtex4-1}
\usepackage{graphicx}
\usepackage{dcolumn}
\usepackage{bm}
\usepackage{color}

\newcommand{\aap}{A\&A}
\newcommand{\adndt}{At. Data Nucl. Data Tables}

\newcommand{\apjl}{Astrophys. J. Lett.}

\newcommand{\apjrnl}{Astrophys. J.}

\newcommand{\jpgnp}{J. Phys. G: Nucl. Phys.}
\newcommand{\mnras}{Mon. Not. R. Astron. Soc.}
\newcommand{\mpla}{Mod. Phys. Lett. A}
\newcommand{\npa}{Nucl. Phys. A}
\newcommand{\physrep}{Phys. Rep.}
\newcommand{\plb}{Phys. Lett. B}

\begin{document}

\title{Impact of Nuclear Mass Uncertainties on the $r$-process}

\author{D. Martin}
\email[]{dirk.martin@physik.tu-darmstadt.de}

\affiliation{Institut f\"ur Kernphysik, Technische Universit\"at Darmstadt, Schlossgartenstr. 2,
Darmstadt D-64289, Germany}
\affiliation{GSI Helmholtzzentrum f\"ur Schwerionenforschung GmbH, Planckstrasse 1, Darmstadt D-64291, Germany}

\author{A. Arcones}
\email[]{almudena.arcones@physik.tu-darmstadt.de}

\affiliation{Institut f\"ur Kernphysik, Technische Universit\"at Darmstadt, Schlossgartenstr. 2,
Darmstadt D-64289, Germany}
\affiliation{GSI Helmholtzzentrum f\"ur Schwerionenforschung GmbH, Planckstrasse 1, Darmstadt D-64291, Germany}

\author{W. Nazarewicz}
\email[]{witek@frib.msu.edu}
\affiliation{Department of Physics and Astronomy and FRIB Laboratory,
Michigan State University, East Lansing, Michigan 48824, USA}
\affiliation{Institute of Theoretical Physics, Faculty of Physics, 
University of Warsaw, 02-093 Warsaw, Poland}

\author{E. Olsen}
\email[]{olsene@nscl.msu.edu}
\affiliation{NSCL/FRIB Laboratory, Michigan State University, East Lansing, Michigan 48824, USA}

\date{\today}

\begin{abstract}
Nuclear masses play a fundamental role in understanding how the 
heaviest elements in the Universe are created in the $r$-process.
We predict $r$-process nucleosynthesis yields using neutron capture 
and photodissociation rates that are based on the nuclear density functional
theory. Using six Skyrme energy density functionals based on different optimization protocols, 
we determine for the first time systematic uncertainty bands -- related to mass modeling -- for $r$-process abundances
in realistic astrophysical scenarios. We find that features of the
underlying microphysics make an imprint on abundances especially in the 
vicinity of neutron shell closures: Abundance
peaks and troughs are reflected in trends of neutron 
separation energy. Further advances in the nuclear theory  
  and experiments, when linked to
observations, will help in the understanding of astrophysical conditions in extreme
$r$-process sites.
\end{abstract}


\maketitle

{\it Introduction} --- 
Understanding the origin of elements in nature is one of the outstanding questions in science.
Here, the synthesis of the heavy elements represents a difficult interdisciplinary challenge. 
Half
of the heavy elements up to bismuth and all of the thorium and uranium in the Universe
are produced by the rapid capture of neutrons in the $r$-process \cite{Arnould200797}. This process requires high neutron densities and involves
extreme neutron-rich nuclei not yet produced in the laboratory. 

In recent years, much progress has been made toward solving this problem in
both astrophysics and nuclear physics. In astrophysics,
multidimensional hydrodynamic simulations including improved
microphysics indicate that (i) neutrino-driven winds following
core-collapse supernovae are not neutron-rich enough to produce heavy
elements as suggested in \cite{Woosley.etal:1994} (see
Ref.~\cite{Arcones.Thielemann:2013} for a review), (ii) a rare kind of
core-collapse supernova triggered by magnetic fields leads to
neutron-rich jets where the $r$-process can occur
\cite{Winteler.etal:2012, Moesta.etal:2014, Nishimura.etal:2015}, and
(iii) neutron star mergers (as suggested in
\cite{Lattimer.Schramm:1974} and preliminarily studied in
\cite{Freiburghaus.etal:1999a}) -- are excellent candidates for the
synthesis of heavy elements
\cite{Korobkin.etal:2012,Goriely.etal:2011,Wanajo.etal:2014} even if
their contribution to the early Galaxy is still under discussion. 
  Some studies show that neutron star mergers provide an important
  contribution to the Solar System $r$-process, but this is not enough to
  explain the abundances in the oldest observed stars; see, e.g.,
  \cite{Matteucci.etal:2014,Wehmeyer.etal:2015,Shen.etal:2015}. In
  contrast, other models can explain the $r$-process abundances at all
  metallicities solely based on the neutron star merger 
  scenario \cite{Ishimaru.etal:2015}.

Experimentally, there has been impressive progress in approaching
 $r$-process nuclei; see \cite{Hosmer.etal:2010,Madurga.etal:2012,Watanabe.etal:2013,Lorusso.etal:2015,Atanasov15,Mazzocchi15}, and references quoted therein. New-generation radioactive ion beam facilities
\cite{Decadal2012,NUPECCLRP2010,NSACLRP2015} will be able to reach a
range of nuclei never possible before, including the neutron-rich
frontier of the nuclear landscape. 
Theoretically, there have been major advances in both nuclear modeling and astrophysical
simulations, greatly facilitated by high-performance computing
\cite{ModPhysLettA.29.1430010,UNEDF,BinderCC,NucEFT,Hagen48Ca,Hebeler.etal:2015,JankaSN,Couch15,CITA2015,Mosta2015}. When it comes to the nuclear input for the $r$-process, one relies on global predictions of nuclear properties \cite{Arcones.MartinezPinedo:2011,Goriely.etal:2013,Mumpower2015,Mendoza-Temis.etal:2015}. A microscopic tool that is well suited
to provide quantified microphysics anywhere on the nuclear chart is the nuclear density functional theory (DFT) \cite{RevModPhys.75.121}
based on a realistic energy density functional (EDF) representing the density-dependent effective nuclear interaction. This approach is capable of predicting a variety of observables needed and is able to assess the uncertainties on those observables, both statistical and systematic \cite{ThErrors,DFTBayes}.
Such a capability is essential in the context of making extrapolations into the regions where experiments are impossible \cite{Erler.etal:2012,gao2013,Kort13,DFTBayes}.

{\it Objectives} --- 
In this study, we present the impact of systematic uncertainties 
on nuclear masses for
the two
most promising astrophysical $r$-process scenarios: neutron star mergers
\cite{Korobkin.etal:2012} and jetlike supernovae
\cite{Winteler.etal:2012}. 
The systematic (model) uncertainties are estimated 
by considering several EDFs optimized to experimental data. The corresponding systematic error thus represents the root-mean-squared spread
of predictions of different EDF parametrizations obtained by means
of diverse fitting protocols. In the absence of the exact
reference model, such an intermodel deviation should be viewed as a
rough approximation to the systematic error. A similar strategy was employed to estimate the position of neutron and proton drip lines \cite{Erler.etal:2012,Afanasjev13}, to study the landscape of two-proton radioactivity \cite{Olsen13,*Olsen13a}, and to assess neutron-skin uncertainties \cite{Kort13}.
This approach is complementary to varying
individual masses within some assumed error bars \cite{Mumpower2015,Mumpower2015a,Mumpower2016}
or considering mass models and mass formulas based on vastly different physical
approaches \cite{Farouqi.etal:2009, Arcones.MartinezPinedo:2011}. 

{\it Method} --- 
Nucleosynthesis calculations are performed within a complete nuclear
reaction network (see Ref.~\cite{Martin.etal:2015}
and references therein). The masses of even-even nuclei were computed in Ref.~\cite{Erler.etal:2012} for six different Skyrme EDFs: SkM$^*$ \cite{NucPhysA.386.79}, SkP \cite{NucPhysA.422.103}, SLy4 \cite{NucPhysA.635.231}, SV-min \cite{PhysRevC.79.034310}, UNEDF0 \cite{PhysRevC.82.024313}, and UNEDF1 \cite{PhysRevC.85.024304}. (For the corresponding mass tables, see Ref.~\cite{massexplorer}.) 
The masses of odd-$A$ and odd-odd isotopes were obtained by 
adding computed average pairing gaps to the binding energy of the corresponding zero-quasiparticle vacuum obtained by averaging binding energies of even-even neighbors; see supplemental material of Ref.~\cite{Erler.etal:2012}.

For each EDF model, we compute Maxwellian-averaged $({\rm n}, \gamma)$ reaction rates in the framework of the statistical model \cite{Holmes.etal:1976} using the TALYS code \cite{Talys1.6} with standard input (apart from the masses). This model relies on the assumption of a thermodynamic equilibrium in combination with compound nucleus reactions for excited states.
Photodissociation rates are determined by a detailed balance with partition functions that are consistently obtained from the statistical model. In the reaction network, all $(\gamma, {\rm n})$ and $({\rm n}, \gamma)$ 
rates are replaced by new ones based on individual EDF 
parametrizations. 
Beta decay and fission rates have been taken at default values; those will be addressed
in forthcoming work.
Beta decay rates are not expected to dramatically change
when varying mass models as compared to
  photodissociation reaction rates, which depend exponentially on the
  separation energy. Moreover, the impact of beta decays on the
  abundances have been shown to be moderate~\cite{Hosmer.etal:2010,Nishimura.etal:2012,Madurga.etal:2012,Lorusso.etal:2015,Eichler.etal:2015}.
For fission we use the same input as in Refs.~\cite{Korobkin.etal:2012,Martin.etal:2015} that
  is based on Refs.~\cite{Panov.etal:2001,Panov.etal:2005,Panov.etal:2010}. 
  Therefore, fission barriers and yield distributions are
  not consistent with the underlying mass model. 
 Note, however,  that the majority of models of fission yields currently used in
$r$-process simulations are based on a simplistic barrier penetration
approach that employs a notion of the static fission barrier. As this approach ignores collective dynamics,  current fission models are prone to errors that exceed uncertainties related to the assumed input
(mass models); see the discussion in Ref. \cite{Sadhukhan.etal:2016}.
  In neutron star
  mergers, fission can have a big impact on the abundances around the
  second peak \cite{Korobkin.etal:2012,Goriely.etal:2013,Eichler.etal:2015,Shibagaki.etal:2016}. 
  Consequently, in the results presented below, the impact of mass uncertainties on this 
  region should be taken with caution, while for the third $r$-process peak our
  conclusions are robust. 

\begin{figure}[!htb]
 \includegraphics[width=1.0\linewidth]{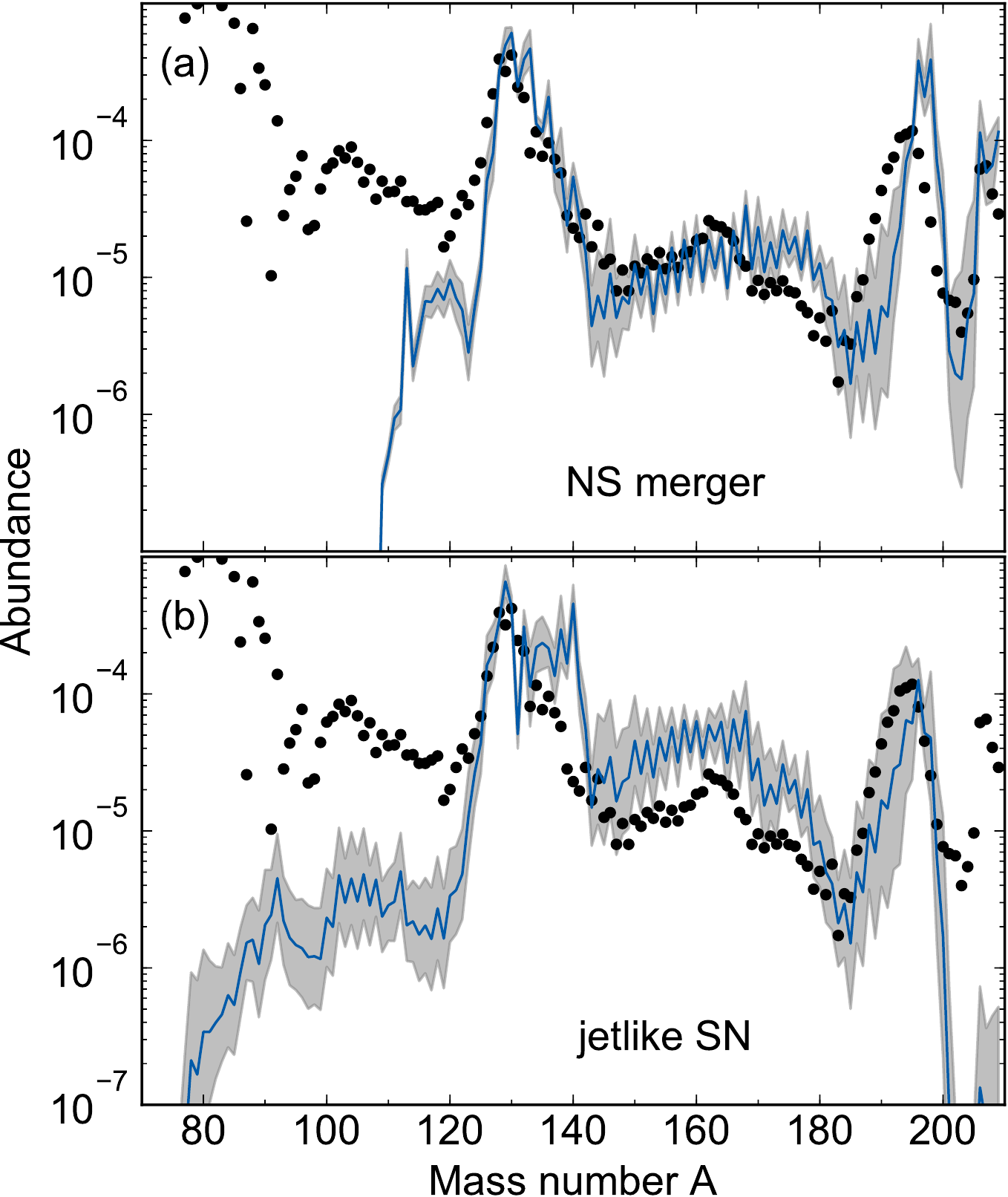}
\caption{(Color online) Predicted abundance distributions 
for neutron star mergers (top) and jetlike supernovae (bottom). Dots indicate the Solar System $r$-process abundances. The systematic uncertainties (gray bands) are due to variations
 of the masses predicted in the six Skyrme-DFT models of Ref.~\cite{Erler.etal:2012}. The mean predicted abundances are marked by the solid line.}
\label{fig:ab_band}
\end{figure}
{\it Results} --- 
The neutron capture and photodissociation rates based on Skyrme-DFT masses \cite{Erler.etal:2012}
 have been used to calculate $r$-process abundances in neutron star
mergers and jetlike supernovae. The various EDFs lead
to different abundances; this  variation is expected given their different optimization schemes. Therefore, when the six mass sets are
considered, we obtain a systematic uncertainty band for the $r$-process abundances
as shown in Fig.~\ref{fig:ab_band}.

The solar system $r$-process
abundances do not always lie within the uncertainty band. This 
indicates that improvements in nuclear
physics and astrophysical inputs are still necessary. However, important hints for future
developments are offered by inspecting our uncertainty estimates. For
example, the uncertainty band is not uniform but strongly depends on
the mass number. This  is in contrast to the sensitivity studies
summarized in Ref.~\cite{Mumpower2016}. Therein, they find a broad and homogeneous
uncertainty band for all mass numbers as a consequence of randomly
varying individual masses within the same range. In our work, mass variations are correlated 
through the microscopic framework employed. 
The mass dependence can be seen in the abundances for neutron star mergers in Fig.~\ref{fig:ab_band}(a) where the uncertainty band is narrow for the
second $r$-process peak ($A\sim 130$) and broadens up before the third
peak ($A\sim 195$). The second peak gets its major contribution from
fission \cite{Korobkin.etal:2012, Eichler.etal:2015}. Since in this pilot study we are using the same fission barrier and
yield distribution data for the six mass sets, only small
variations are expected in this region. In contrast, the evolution of
nuclear masses as a function of the neutron number critically impacts the abundances around shell closures when
nuclei change character from deformed to spherical and back to deformed again. This occurs around neutron
magic numbers where the abundance peaks form, thus leading to a broad
uncertainty band. Since fission plays a minor role for  jetlike supernovae in Fig.~\ref{fig:ab_band}(b), the uncertainty band is broader than in the case of neutron star mergers before and after the
second $r$-process peak.

\begin{figure*}[!htb]
 \includegraphics[width=0.95\linewidth]{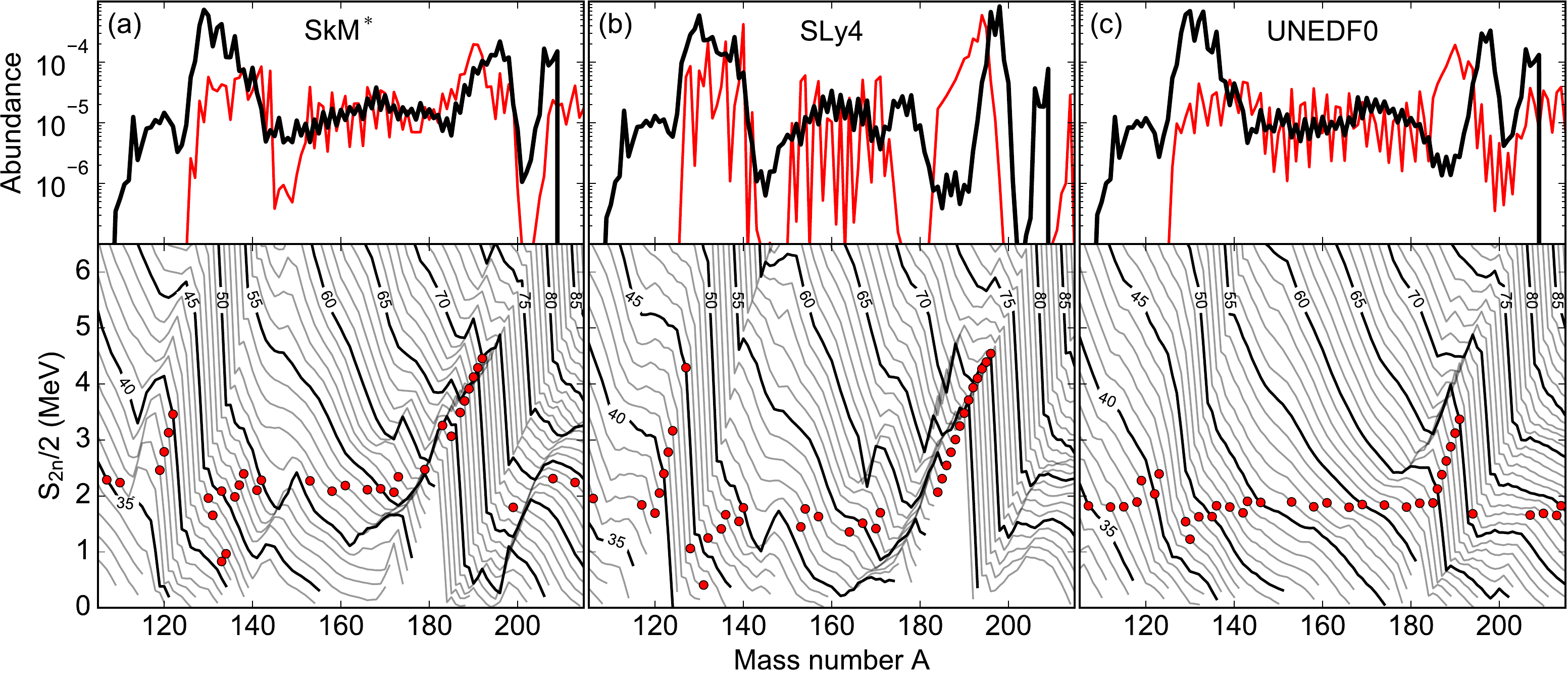}
\caption{(Color online) Predictions of the SkM$^*$, SLy4, and UNEDF0
models.
Bottom: (Half of the) two-neutron separation energies along different isotopic chains (gray lines); every fifth isotope chain is plotted with black lines. The red dots correspond to the freeze-out abundances, i.e., the $r$-process path before matter starts to decay to stability. A few 
 dots around $A \sim 135$ lie below the separation energy 
 that is expected from 
 $({\rm n},\gamma)-(\gamma,{\rm n})$ equilibrium. Here, fission of 
 nuclei with $A>240$ populates regions beyond the r-process path. 
Top: The freeze-out (thin red lines) and final 
 (thick black lines) abundances for a neutron star merger scenario.}
\label{fig:ab_s2n}
\end{figure*}

In order to better understand the impact of nuclear masses on the predicted abundances, in Fig.~\ref{fig:ab_s2n} (bottom panels)
we analyze the trend of two-neutron separation energies $S_{2{\rm n}}$ for the
 SkM$^*$, SLy4, and UNEDF0
models. The red dots indicate the $r$-process path at freeze-out,
i.e., the mass number with the highest abundance for every isotopic
chain. At $r$-process freeze-out, most of the
neutrons are consumed, and neutron-rich material starts to decay to
stability. The abundances at freeze-out are marked by thin red lines in the upper panels, while
the final abundances, following beta decays back to the valley of stability, are indicated by 
thick black lines.

The most remarkable features are
the rapid variations of separation energies of neutron-rich nuclei at $A \approx 120-140$ and $A \approx 180-200$
associated with the neutron magic numbers $N=82$ and $126$,
respectively. Moreover, one can identify the formation of peaks shown
in the freeze-out abundances with the regions where matter accumulates
(red dots). The third $r$-process peak at freeze-out
is located at smaller mass numbers than in the final abundances, pointing to some
important reactions occurring during the decay to stability. Beta
decays keep the mass number constant or reduce it in the case of
beta-delayed neutron emission (which can be significant as it increases the number of
neutrons). This deviation indicates that the
shift of the third peak is due to neutron captures
\cite{Surman.Engel:2001, Arcones.MartinezPinedo:2011}.
Neutrons are thus critical in understanding both the evolution toward
stability and the final abundances. In addition to the few leftover
neutrons after freeze-out, there are also contributions from
beta-delayed neutron emission and fission. In the case of neutron star
mergers \cite{Eichler.etal:2015}, there are many neutrons produced in
fission and this leads
to a more pronounced shift of the third peak than in jetlike supernovae
(Fig.~\ref{fig:ab_band}).

The results shown in Fig.~\ref{fig:ab_s2n} were obtained using three EDFs developed
using different optimization protocols. The functional SkM$^*$ \cite{NucPhysA.386.79} is the traditional Skyrme EDF fitted to binding energies and charge radii of selected spherical nuclei, spin-orbit splitting in $^{16}$O, giant resonance energies in $^{208}$Pb, and fission barriers. 
The functional SLy4 \cite{NucPhysA.635.231} was optimized with a focus on neutron-rich
nuclei and neutron matter. In addition to properties of spherical nuclei, this functional was also constrained to basic properties of  symmetric nuclear matter and the equation of state for pure neutron matter. Finally, a more recent model UNEDF0 \cite{PhysRevC.82.024313} was carefully 
optimized to a large database including masses of spherical and deformed nuclei, charge radii of spherical nuclei, odd-even mass differences, and selected nuclear matter properties.
These models have different performances when it comes to masses. The older models SkM$^*$ and SLy4 
yield a large rms deviation from experiment, around and greater than
5\,MeV. This can be attributed to an overemphasis on doubly magic nuclei during optimization as well as a fairly limited data set. As discussed in Ref.~\cite{PhysRevC.82.024313}, the functional UNEDF0, with its rms deviation of 1.45 MeV, is probably within a few hundreds of keV of a globally optimal mass fit within the Skyrme EDF parameter space.
To put things in perspective, we note that the best overall
agreement with experimental masses, obtained with the Skyrme
EDF, is around 600\,keV \cite{GorielyMass}. However, this excellent result was
obtained at a price of several phenomenological corrections on top of the original Skyrme-DFT model.

\begin{figure}[!htb]
 \includegraphics[width=1.0\columnwidth]{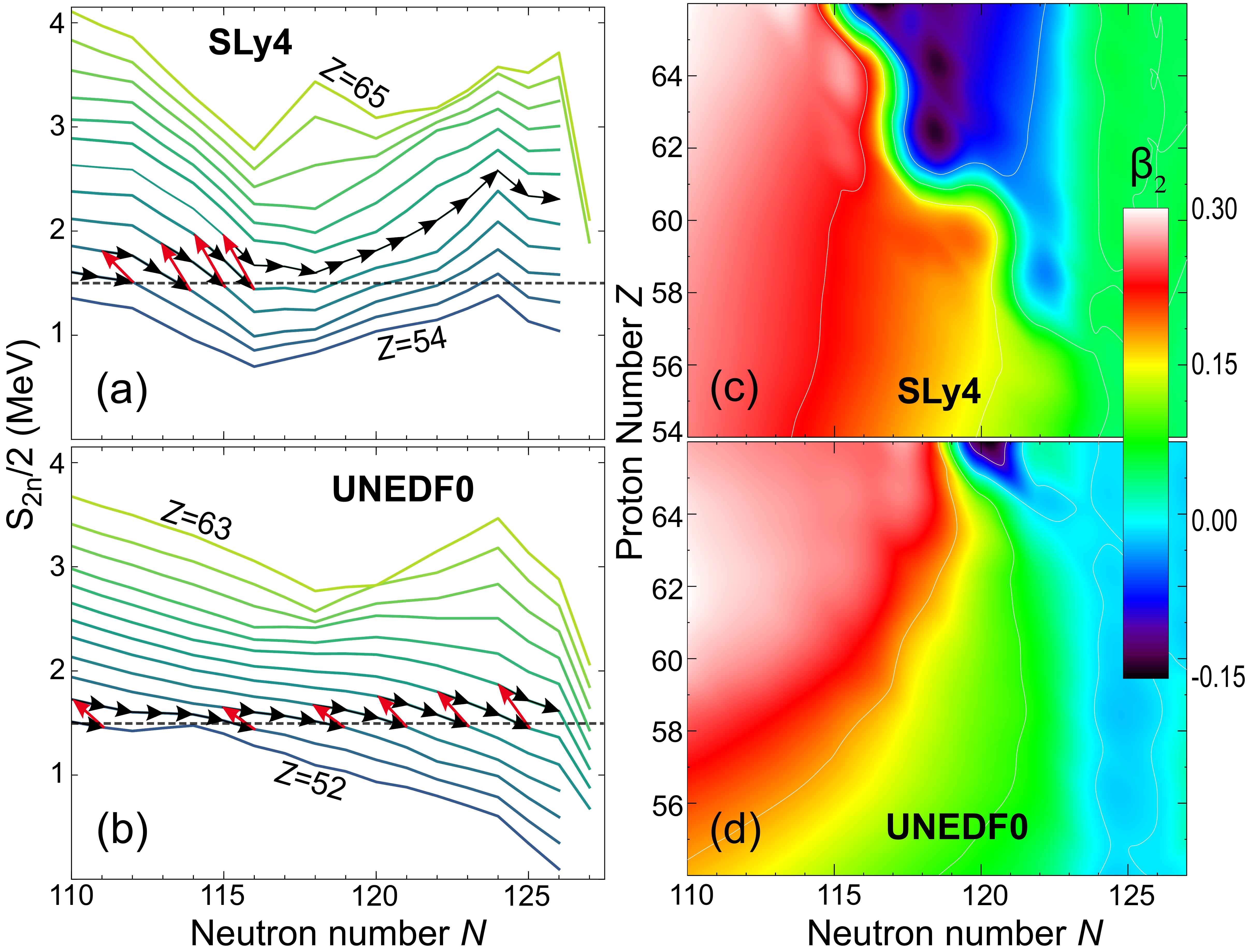}
\caption{(Color online) Left: Half of the two-neutron
    separation energies predicted by the models SLy4 (a) and UNEDF0
  (b) versus the neutron number in the region where the third $r$-process
  peak forms. The dashed line marks the approximate $r$-process path during $({\rm n},\gamma)-(\gamma,{\rm n})$ equilibrium ($S_{2{\rm n}}/2 = 1.5$~MeV). Black arrows indicate
neutron captures; red arrows mark beta decays along the path.
Right: The corresponding quadrupole deformations $\beta_2$ for SLy4 (c) and UNEDF0 (d).}
\label{fig:s2n_126}
\end{figure}

The parametrization SkM$^*$ is the one that
leads to the smallest shift of the third peak from freeze-out to final
abundances. For SLy4, the third peak is
strongly shifted compared to SkM$^*$, and there is a big trough in abundances before
it. The trough corresponds to the region without dots in
 the lower panel of Fig.~\ref{fig:ab_s2n}(b) ($A \sim 180$). 
The origin of this trough is the nonmonotonic behavior of the separation 
energy \cite{Arcones.MartinezPinedo:2011,Arcones.Bertsch:2012} predicted in this 
model. In order to illustrate the behavior of the freeze-out abundances before $N=126$, 
the left panel of Fig.~\ref{fig:s2n_126} shows $S_{2{\rm n}}$ in SLy4 and UNEDF0
in the region where the third $r$-process peak forms. Far from stability, the 
$r$-process path stays at an almost constant neutron separation energy 
during the $({\rm n},\gamma)-(\gamma,{\rm n})$ equilibrium. This value 
is marked in Figs.~\ref{fig:s2n_126}(a) and \ref{fig:s2n_126}(b) by the
dashed line at a representative value of $S_{2{\rm n}}/2 = 1.5$\,MeV. Note 
that the explanation here is valid for any $({\rm n},\gamma)-(\gamma,{\rm n})$ 
equilibrium in general, but inspired by the values found
in Fig.~\ref{fig:ab_s2n}. When matter approaches this limit, it
stays there until a beta decay occurs. Afterwards, more neutron
captures are possible until the separation energy reaches this limit again. Figure~\ref{fig:s2n_126}(a)
shows that for SLy4 the nonmonotonic behavior of $S_{2{\rm n}}$ leads to a long sequence of neutron captures at a fixed $Z$ value; hence, it results in a trough in the abundances versus mass number.
In the case of UNEDF0, on the other hand, after a beta
  decay, the $({\rm n},\gamma)-(\gamma,{\rm n})$-equilibrium $S_{\rm
    n}$ is reached again after only a few neutron captures. The nonmonotonic behavior of neutron separation energies can be traced back to
 structural changes due to rapid shape transitions from prolate to spherical to oblate. While the global deformation patterns predicted
by various EDFs are fairly similar \cite{Erler.etal:2012}, the subtle
details of shape transitions are predicted differently. This is illustrated in Figs.~\ref{fig:s2n_126}(c) and \ref{fig:s2n_126}(d) for SLy4 and UNEDF0, respectively.

The functional UNEDF0, informed by experimental masses in spherical and deformed nuclei, 
exhibits a smoother evolution of $S_{2{\rm n}}$ versus $A$ than
other functionals considered. In fact, it is so much less steep for $N=82$ that the shell closure is quenched for very neutron-rich nuclei with $Z\le 40$; see Fig.~\ref{fig:ab_s2n}(c). This results in the lack of the second peak in the freeze-out abundances. However, as discussed above, the second peak in the final abundances has its origin in fission.

The freeze-out abundances and the evolution toward stability to
produce the final abundances depend on astrophysical
conditions. For the jetlike supernova trajectory the freeze-out path
is closer to stability (i.e., at higher neutron separation energies) and there are less
neutrons available during the decay because of the reduced effect of fission. The
features that we have explained before for the neutron star merger
affect the jetlike supernova abundance differently. Therefore, reducing
uncertainties in the nuclear physics input should enable us to use
observations to constrain and understand the astrophysical conditions
related to the $r$-process site. 

{\it Conclusions} --- In summary, we have shown that detailed features of nuclear mass evolution toward the
neutron drip line are critical to understand both the abundances and the origin of
heavy elements in the $r$-process. Of utmost importance are the regions around magic numbers where separation energies vary rapidly due to spherical shell closures and shape changes. The systematic uncertainty band obtained within the deformed Skyrme-DFT approach exhibits significant variations with particle number. It is encouraging, however, that in certain mass regions the model error is fairly small; i.e., the intermodel consistency of our results can be high. 
The regions characterized by the broad uncertainty bands in Fig.~\ref{fig:ab_band} are indicative of model differences far from stability, where the theory relies on (sometimes extreme) extrapolations. To reduce the uncertainties, the development of nuclear EDFs of spectroscopic quality, constrained by data on the most neutron-rich nuclei reachable in experiments, is needed. In this respect, the $r$-process abundance predictions presented in this work aim at
 assessing the current status of theoretical mass modeling at the limits
of nuclear binding and also
provide a useful benchmark for future improvements. In the next step,
we intend to improve other microphysics input, such as fission yields
and beta decay rates, as well as to explore additional astrophysical environments. 

It would be interesting to evaluate systematic uncertainty bands for
other microscopic mass models based on effective interactions (or
EDFs). In this context, we note that mass predictions performed within the 
covariant DFT \cite{Afanasjev13} have provided separation-energy
uncertainties remarkably similar to those from Skyrme-DFT
\cite{Erler.etal:2012}. Finally, we wish to emphasize that our
methodology, based on model-based extrapolations (hence, considering
correlations between predicted masses) is complementary
  to and different from sensitivity studies based on individual mass variations
(see e.g., \cite{Mumpower2016}) resulting in a homogeneous uncertainty band
for all masses.

\begin{acknowledgments}
We thank Julia Bliss for useful discussions about the nucleosynthesis. This work was supported by Helmholtz-University Young Investigator
Grant No. VH-NG-825; by the BMBF under Grant No. 05P15RDFN1; by the U.S. Department
of Energy, Office of Science, under Awards
No.~\mbox{DE-SC}0013365~(Michigan State University), No.~\mbox{DE-SC}0008511~(NUCLEI SciDAC-3 Collaboration), and No.~\mbox{DE-NA}0002574~(the Stewardship Science Academic Alliances
program).

\end{acknowledgments}

\bibliographystyle{apsrev4-1} 
\bibliography{bibliography}

\end{document}